%% file: main_6page.tex
\documentclass[letterpaper, 10 pt, conference]{ieeeconf}  

\usepackage{graphicx}
\usepackage{amsmath}
\usepackage{amssymb}
\usepackage{booktabs}
\usepackage{caption}
\usepackage{cite}
\usepackage{pifont}
\DeclareMathOperator{\atantwo}{atan2}
\newcommand{\xmark}{\ding{55}}%
\usepackage{xcolor}
\usepackage{multirow}
\usepackage{siunitx}
\SetSymbolFont{operators}{bold}{OT1}{cmr}{b}{n}
\usepackage[breaklinks,colorlinks,bookmarks=false]{hyperref}

\IEEEoverridecommandlockouts                              

\renewcommand{\baselinestretch}{0.95}

\title{\LARGE \bf
RNR-Nav: A Real-World Visual Navigation System \\ Using Renderable Neural Radiance Maps
}

\author{Minsoo Kim$^1$, Obin Kwon$^2$, Howoong Jun$^{1,3}$, and Songhwai Oh$^{1,2,3}$%
\thanks{$^{1}$Interdisciplinary Program in Artificial Intelligence (IPAI) and ASRI, Seoul, Republic of Korea (e-mail: minsoo.kim@rllab.snu.ac.kr, howoong.jun@rllab.snu.ac.kr, songhwai@snu.ac.kr).}
\thanks{$^{2}$Department of Electrical and Computer Enginnering (ECE) and ASRI, Seoul National University, Seoul, Republic of Korea (e-mail: obin.kwon@rllab.snu.ac.kr).}
\thanks{$^{3}$Sequor Robotics, Inc., Seoul, Republic of Korea.}
\thanks{This work was supported by Institute of Information \& Communications Technology Planning \& Evaluation (IITP) grant funded by the Korea government (MSIT) (No. 2019-0-01190, [SW Star Lab] Robot Learning: Efficient, Safe, and Socially-Acceptable Machine Learning).}
\thanks{\textit{(Corresponding author: Songhwai Oh.)}}
}

\begin{document}

\maketitle

\begin{abstract}

\input{abstract/abs2}

\end{abstract}

\section{Introduction}
The development and deployment of visual navigation systems represent a cornerstone in advancing autonomous navigation capabilities.
The ability to understand visual information in environments is essential for intelligent agents to perform visual navigation tasks effectively.
Numerous studies have attempted to represent the visual information from the observed scenes in various forms, including the adoption of grid-form maps \cite{grid_1,grid_3,grid_4}  where each grid cell represents the environmental information of a corresponding region.

Among the myriad of approaches, the concept of renderable neural radiance map (RNR-Map) \cite{RNR-Map} has been identified as a promising approach for enhancing scene representation with visual information in real-time procedure.
RNR-Map is a grid-form map that contains visual information as latent codes at each grid cell, which can be rendered to image signals by neural radiance fields (NeRF)  techniques \cite{nerf, gsn}.
%
%
%
The original study \cite{RNR-Map} introduces several applications of RNR-Map for vision-based robotic tasks such as localization and navigation.
%
Though RNR-Map shows significant potential in simulated environments \cite{gibson, mp3d}, real-world environments impose unexplored challenges.
In this paper, we address such challenges encountered in real-world deployment. 

\begin{figure}[t!]
{\centering
\includegraphics[width=0.9\linewidth]{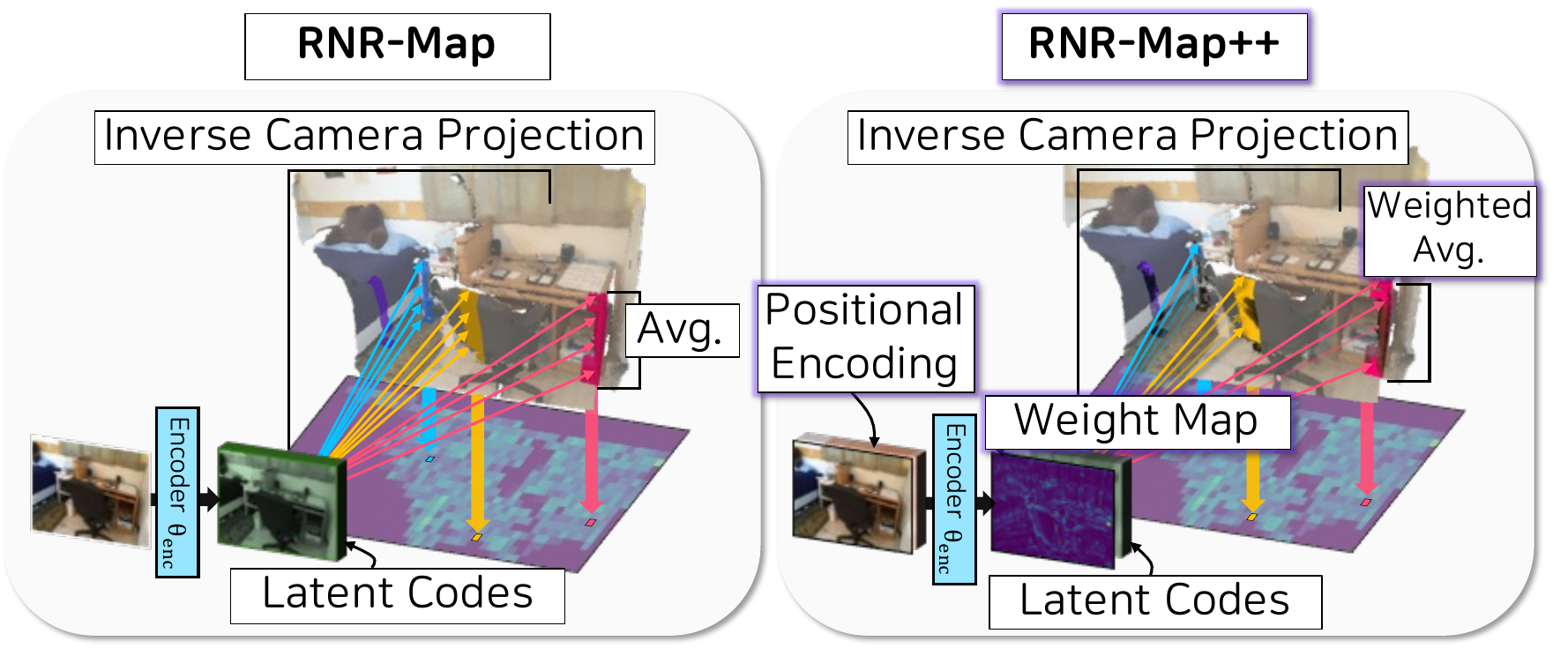}
}\centering
\vspace{-0.2cm}
\caption{\small\textbf{Comparison between the proposed RNR-Map++ and the original RNR-Map.}}
\label{fig:map}
\vspace{-0.7cm}
\end{figure}

%
%
The original RNR-Map shows poor image reconstruction results for the real-world dataset due to the loss of visual information that occurs during the mapping procedure. 
%
%
The projection simply averages multiple latent codes into a single vector for each grid, which blends distinct visual and spatial details.
To achieve environment-agnostic deployment, we initially focus on improving image reconstruction quality by embedding the visual information more effectively.
Our advanced version of RNR-Map, RNR-Map++ shown in Figure \ref{fig:map}, implicitly learns the importance of each pixel and leverages such values before aggregating to a single vector.
RNR-Map++ also adopts positional encoding of 3D positions of each image pixel to contain spatial information.

For the next step, we develop a robust and fast localization process using RNR-Map++.
While \cite{RNR-Map} offers a fast localization method based on cross-correlation operation, inaccuracies in RGB-D images from real-world environments led to unstable predictions that are far from the actual pose.
%
%
%
The RNR-Map suggests a supplementary method based on optimizing the photometric loss, utilizing noisy odometry information.
%
However, it requires rendering as well as optimization processes for several iterations, which consumes large computation time.
%
%
Here, we propose a novel localization method using RNR-Map++ by employing the classical particle filter algorithm \cite{particle} to stabilize the pose estimation.
%
%
%
Our experiment results demonstrate that the particle filter considerably improves localization accuracy.
Moreover, our localization speed appears much faster than the optimization \cite{RNR-Map} or rendering-based \cite{locnerf} localization approaches.

Building on these advancements, we propose RNR-Nav, a novel visual navigation framework for a real-world robot system utilizing RNR-Map++. 
%
%
With the given RNR-Map++ and occupancy map of the environment, our RNR-Nav system is able to find the target pose of the query image and navigate to the goal point by localizing the current pose and planning a path to the goal.
In the image-goal navigation experiment in the real-world environment, the proposed framework achieves a success rate of 84.4\%.

Our main contributions can be summarized as follows:

\begin{itemize}

\item The proposed method, RNR-Map++, mitigates information loss by estimating the importance of each pixel using a weighted map and positional encoding.
\item We propose a novel localization framework using RNR-Map++, alleviating unstable localization in the real world by adopting the particle filter.
\item The extensive experiments show the effectiveness of the proposed RNR-Map++ and navigation framework, RNR-Nav, even in noisy real-world environments.

\end{itemize}

\section{Related Work}
%
%
With the advancement of NeRF \cite{nerf} techniques, several localization \cite{locnerf, nerfloc} and mapping \cite{coslam, nerf_slam} approaches with NeRF have been developed.
Such methods utilize a photometric loss from the rendered image and the observed image information to estimate the camera pose and learn map representation.
NeRF itself can be regarded as a scene representation that contains visual information about any specific scene.
Loc-NeRF \cite{locnerf} proposed a rendering-based localization method combining particle filter and NeRF.
This method calculates the weight of each particle by rendering the image value from the NeRF representation and comparing it with the image observation.
However, this dependency on the rendering process introduces considerable time and memory allocation, as well as extensive optimization time for training NeRF representation on a scene-by-scene basis.
In contrast, our proposed localization framework does not require rendering to calculate particle weights by utilizing cross-correlation operations.
Furthermore, most of the existing methods are limited to simulated environments or localization experiments in small real-world environments.
The introduction of RNR-Map++ enables immediate real-world application without a need for fine-tuning, offering flexibility for deployment across diverse environments.

There have been similar real-world navigation approaches with informative scene representations \cite{nlmap, vlmaps, usa, clipfields} other than renderable information. 
They use pretrained features from vision-language models for the semantic scene interpretation.
However, those approaches prioritize high-level semantic policy execution, often sidelining robot-centric aspects like localization or path planning.
Our proposed method emphasizes the active utilization of embedded information for precise robot localization as well as navigation to the target place specified by images.
%
%


\section{Preliminary}
In this section, we define some notations and review the concept and details of the RNR-Map \cite{RNR-Map}.

\subsection{Map Construction and Rendering}
 
The RNR-Map is made from RGB-D images of the desired environment. In addition, the poses and intrinsic parameters of the camera have to be known. Those images are fed sequentially to the encoder, which consists of convolutional layers. When the $t$-th RGB-D image $I_{t} \in \mathbb{R}^{H \times W \times 4}$ is given as an input, the encoder encodes the image into latent codes in identical size, $C_{t} \in \mathbb{R}^{H \times W \times D}$. Here, $H$, $W$, and $D$ refer to the height, width of the image, and the dimension of the latent codes, respectively.
For each pixel $(h,w)$, we can calculate 3D position $\mathbf{q}_{(h,w)}=(q_{(h,w),x},q_{(h,w),y},q_{(h,w),z})$ in world coordinate system of each pixel by using known camera intrinsic parameters, the depth value, and the pose of the image $p_{t} = [\mathbf{R}|\mathbf{t}]$. 
Then, the position is discretized into the grid with position $(u,v)$ in the map as follows:


%

\vspace{-0.2cm}
\begin{equation}
    (u,v) = \left ( \left \lfloor \frac{q_{(h,w),x}}{s} \right \rfloor, \left \lfloor \frac{q_{(h,w),y}}{s} \right \rfloor \right),
\end{equation}
\noindent
where the grid size $s$ denotes the height and length of each grid in RNR-Map. The projection process averages latent codes $c_{(h,w)} \in \mathbb{R}^{D}$ in $C_{t}$ belong to the same grid $(u,v)$ into a single latent code. Finally, the map $\mathbf{M} \in \mathbb{R}^{U \times V \times D}$ is constructed with height $U$ and width $V$. In addition, the RNR-Map keeps track of the numbers of the averaged vectors in each grid as a mask $\mathbf{N} \in \mathbb{R}^{U \times V}$. Then, each component at $(u,v)$ of the map and the mask, $M_{(u,v)} \in \mathbb{R}^{D}$ in $\mathbf{M}$ and $N_{(u,v)} \in \mathbb{R}$ in $\mathbf{N}$ respectively, can be derived as follows:

\vspace{-0.4cm}
\begin{align} 
\label{projection}
\begin{split}
   \hspace{-1.0em} P_{(u,v)} &= \left \{ (h,w) \Big| \left ( \left \lfloor \frac{q_{(h,w),x}}{s} \right \rfloor, \left \lfloor \frac{q_{(h,w),y}}{s} \right \rfloor \right) = (u,v) \right \} \\ 
   \hspace{-0.5em} M_{(u,v)} &= \frac{1}{N_{(u,v)}}{\sum_{(h,w) \in P_{(u,v)}} c_{(h,w)}}, \enspace N_{(u,v)} = |P_{(u,v)}|. 
\end{split}
\end{align}
\vspace{-0.3cm}

This process so far with a single image input is named as registration process $F_{reg}$. With RGB-D image $I_{t}$, the pose $p_{t}$, and the parameter of the encoder $\theta_{enc}$, registration process is written as:

\vspace{-0.4cm}
\begin{equation} \label{local_map}
    \mathbf{M}^{l}_{t}, \mathbf{N}^{l}_{t} = F_{reg}(I_{t},p_{t};\theta_{enc}),
\end{equation}
\vspace{-0.5cm}

\noindent
where $l$ denotes that the map represents only local information from a single image. As images are given sequentially, the global map and the global mask at timestep $t$, $\mathbf{M}^{g}_{t}$ and $\mathbf{N}^{g}_{t}$ respectively, should be updated from $\mathbf{M}^{g}_{t-1}$ and $\mathbf{N}^{g}_{t-1}$ using $\mathbf{M}^{l}_{t}$ and $\mathbf{N}^{l}_{t}$.
RNR-Map computes $\mathbf{M}^{g}_{t}$ by weighted averaging both maps utilizing both masks as weights.
%
For simplicity, $\mathbf{M}$ and $\mathbf{N}$ without any superscript refer to the global map and the global mask in the rest of the paper.

The encoder is trained in an autoencoder manner to learn to encode and decode visual information by minimizing photometric loss. Constructed RNR-Map is used as conditioned latent codes to render the image by NeRF \cite{nerf} technique similar to the decoder of GSN \cite{gsn}. The reader may refer to \cite{gsn,RNR-Map} for further information about the decoder.

\subsection{Localization}

RNR-Map \cite{RNR-Map} describes how to use the map for an image-based localization task in two ways: \textit{correlation-based} method and \textit{optimization-based} method. With given key map $\mathbf{M}_{k}$ and a query image $I_{q}$, the localizer figures out which position, \textit{i.e.}, grid cell, on the map is the closest location where the query image is taken. The correlation-based localizer consists of three neural networks, $F_{k}$, $F_{q}$, and $F_{E}$. First, the query map $\mathbf{M}_{q}$ is constructed by (\ref{local_map}) using trained encoder assuming $p=[\mathbf{I}|\mathbf{0}]$. Then, $F_{k}$ and $F_{q}$ having U-Net \cite{unet} architecture convert $\mathbf{M}_{k}$ and $\mathbf{M}_{q}$ into $\mathbf{M}_{k}'$ and $\mathbf{M}_{q}'$, respectively. Since the orientation of the query map is unknown, the localizer expands the query map to 36 maps, $\{ \mathbf{M}_{q,r}' \}_{r=1}^{36}$, by rotating the map 36 discrete angles $\{0^\circ, 10^{\circ}, \cdots, 350^{\circ} \}$. Utilizing $\mathbf{M}_{q,r}'$ as a filter, the localizer conducts convolution with $\mathbf{M}_{k}'$ to gain cross-correlations between the key and the query. Finally, $F_{E}$ transforms the cross-correlations into predicted heatmap $\tilde{\mathbf{H}} \in \mathbb{R}^{U \times V \times R}$, which represents the estimated probability of the pose of the agent on the map. The localizer is trained to minimize the cross-entropy loss between $\tilde{\mathbf{H}}$ and ground truth $\tilde{\mathbf{H}}_{gt}$. The encoder is frozen while training the localizer.






The optimization-based localization method assumes noisy odometry information is given. The inaccurate pose is refined by optimizing the pose to minimize the photometric loss between the observation and the reconstructed image. 

\section{Proposed Method}

%
%
In this section, we elaborate on the motives and specifications of RNR-Map++, our advanced version of RNR-Map, which handles the problems of real-world settings.
Moreover, we describe how we implement the robust localization and navigation system, RNR-Nav, using RNR-Map++ for the real-robot system.

\subsection{Map Construction and Rendering}

\subsubsection{Weighted Map}

Each latent code of the grid cell in the RNR-Map is computed by averaging corresponding latent codes along the height.
However, this approach treats all pixels equally, disregarding the amount of visual information in each region. 
%
Generally, there are relatively less amount of valuable parts with abundant features (\textit{e.g.}, corners and edges) than regions with few features (\textit{e.g.}, flat walls).
Therefore, the grid cell of the RNR-Map originates from a high proportion of uninformative regions.
To rectify this, we adopt a weighted average for the projection.

We modify the encoder to return additional output $B_{t} \in \mathbb{R}^{H \times W}$ along with $C_{t}$.
%
%
The weights $\beta_{(h,w)}$ are calculated by $\beta_{(h,w)} = \{softmax(B_{t})\}_{(h,w)}$.
Each $c_{{(h,w)}} \in C_{t}$ is multiplied by $\beta_{(h,w)}$ to calculate $M_{(u,v)}$, and $N_{(u,v)}$ keeps the sum of the weights rather than the number of the latent codes.
Thus, (\ref{projection}) can be rewritten as follows:

\vspace{-0.5cm}
\begin{align}
\begin{split}
    M_{(u,v)} &= \frac{1}{N_{(u,v)}}\sum_{(h,w) \in P_{(u,v)}}{\beta_{(h,w)} c_{(h,w)}} \\
    N_{(u,v)} &= \sum_{(h,w) \in P_{(u,v)}}{\beta_{(h,w)}}.
\end{split}
\end{align}
\vspace{-0.3cm}

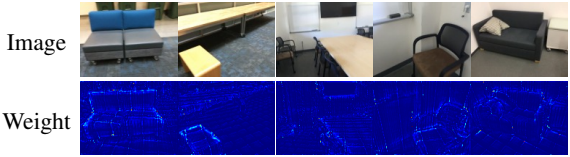
\begin{figure}
\input{figures/weight}
\vspace{-0.2cm}
\caption{\small\textbf{Weights computed by the encoder.} Important regions such as corners or edges have high weight values. Shown real-world images are from ScanNet \cite{scannet} dataset.}
\label{fig:weight}
\vspace{-0.5cm}
\end{figure}

Figure \ref{fig:weight} shows weights inferred by the trained encoder. We can confirm that the high-weight values appear at the corners and edges of each image.

\subsubsection{Positional Encoding}

In an ideal situation such as a simulated setting, the roll and pitch values are fixed to 0, and the movement along the $z$-axis, \textit{i.e.}, vertical axis, is also 0. It enables 3D positions of pixels with identical image coordinate values to be determined solely by distances. Consequently, the encoder can capture the 3D positions of pixels without requiring additional spatial information.
%
%
However, the real-world operation may have rolls, pitches, and $z$-axis displacement.
When encoded latent codes are projected on the 2D RNR-Map, information on the $z$-coordinate can be lost. Therefore, we decide to utilize positional encoding of the world coordinate values of each pixel.
As original NeRF, $\cos$ and $\sin$ functions are applied to 3D positions $\textbf{q}$ after multiplying several frequency values to get positional encoding as follows:

\vspace{-0.5cm}
\begin{equation}
    \hspace{-0.7em} \gamma(\textbf{q}) = \left [ \sin{(\textbf{q})}, \cos{(\textbf{q})}, \cdots, \sin{(2^{L-1} \textbf{q})}, \cos{(2^{L-1} \textbf{q})} \right ]\hspace{-0.3em}.\hspace{-0.5em}
\end{equation}
\vspace{-0.5cm}

\noindent
We concatenate positional encoding vectors to the RGB-D images and hand over them to the encoder as input. Then, (\ref{local_map}) can be rewritten as follows:

\vspace{-0.2cm}
\begin{equation}
    \mathbf{M}^{l}_{t}, \mathbf{N}^{l}_{t} = F_{reg}(I_{t} \oplus \gamma(\textbf{q}) ,p_{t};\theta_{enc}),
\end{equation}
\vspace{-0.5cm}

\noindent
where $\oplus$ denotes the concatenation.




\begin{figure}
\input{figures/recon_quality}
\vspace{-0.3cm}
\caption{\small\textbf{Comparisons of reconstruction qualities of the original RNR-Map and the RNR-Map++.} (a) Images from the ScanNet dataset, which is used for training and validation. (b) Real-world images observed by our robot system, which has different image sizes and camera intrinsics from ScanNet.
Note that the same model trained by ScanNet is used for rendering in both datasets without additional training.}
\vspace{-0.5cm}
\label{fig:recon_quality}
\end{figure}

\subsubsection{Generalizable Input Image Size}

Applying the trained encoder to diverse inputs, especially observations obtained by real-world robots, requires compatibility with varying image scales. To ensure each kernel of convolutional layers gets an identical range of spatial information as input, we standardize the height and width of each pixel on the image plane, \textit{i.e.}, consistent focal length, $f_{x}$ and $f_{y}$. Here, we opt for $128$ as the $f_{x}$ and $f_{y}$ values. It can be satisfied by simply resizing input images utilizing given camera intrinsic parameters.

Figure \ref{fig:recon_quality} depicts the rendering qualities of the original RNR-Map and the advanced RNR-Map++. The proposed RNR-Map++ demonstrates an enhanced capacity to preserve and render finer details within scenes. Furthermore, we can successfully apply the trained encoder and decoder directly to images with different sizes and camera parameters, as Figure \ref{fig:recon_quality}a and \ref{fig:recon_quality}b.
One more notable point is that, in the case of the original RNR-Map, rendering at the same 2D point often appears as a single large chunk. On the contrary, in RNR-Map++, we can observe detailed variations in height at the same location, indicating finer details in rendering. It suggests that with the addition of positional encoding within the same grid, spatial information has been augmented, enabling more precise distinctions during rendering.

\subsection{Localization}

\subsubsection{Correlation-Based Localization}

\begin{figure}[t!]
{\centering
\includegraphics[width=1.0\linewidth]{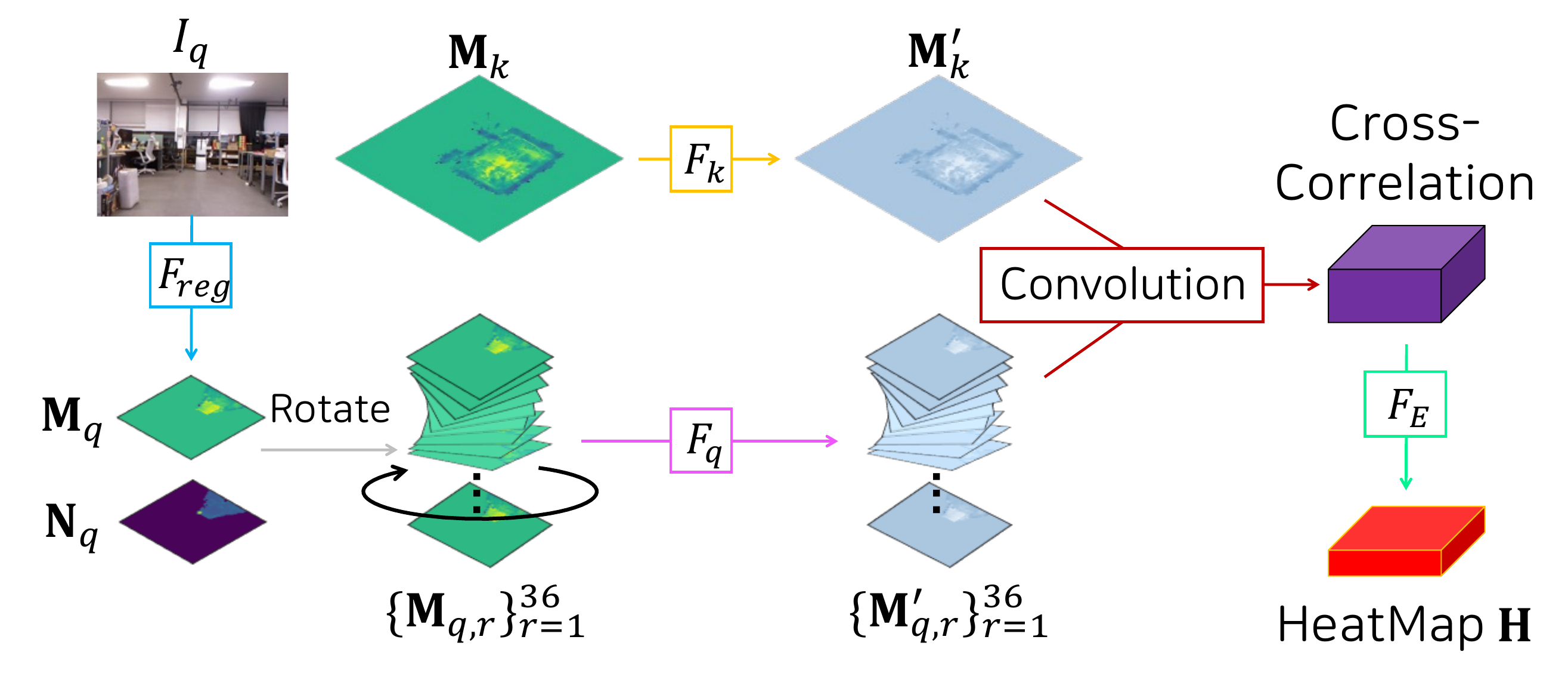}
}\centering
\vspace{-0.7cm}
\caption{\small \textbf{Workflow of the correlation-based localization.} Note that weight mask $\mathbf{N}_{q}$ is used for rotating query map $\mathbf{M}_{q}$.
}
\label{fig:localizer}
\vspace{-0.3cm}
\end{figure}

%
%
Since we adopt the weighted map for constructing RNR-Map++, we revise the way to compute rotated query maps. To maintain the importance of latent codes during rotation, we use the weight mask $\mathbf{N}_{q}$ as well as $\mathbf{M}_{q}$ from (\ref{local_map}). We rotate $\mathbf{M}_{q}$ in 36 discrete angles by interpolation using weight mask $\mathbf{N}_{q}$ as follows:

\vspace{-0.2cm}
\begin{equation}
    \mathbf{M}_{q,r} = \frac{(\mathbf{{M}_{q} \odot \mathbf{N}_{q}})_{r}}{\mathbf{N}_{q,r}}, \quad r \in \{1, \cdots, 36\}.
\end{equation}
\vspace{-0.4cm}

\noindent
where $\odot$ denotes the element-wise product and division is also an element-wise operation. Then, we convert $\mathbf{M}_{q,r}$ to $\mathbf{M}_{q,r}'$ with $F_{q}$. The overall framework is described in Figure \ref{fig:localizer}.
%
%
%
%
The estimated heatmap is handed over to a particle filter.

\subsubsection{Particle Filter}

We propose a robust and real-time applicable localization method with RNR-Map++ adopting particle filter \cite{particle}, assuming noisy odometry information is provided.
%
%
%
%
%
%
For $N_{p}$ particles, each particle has 3-DoF pose value $\{ (u_{i},v_{i},\alpha_{i}) \}_{i=1}^{N_{p}}$: position $(u_{i},v_{i})$ on the grid-map and the orientation $\alpha_{i}$. We use heatmap $\mathbf{H} \in \mathbb{R}^{U \times V \times R}$ of 3-DoF poses as estimated weights of particles after normalizing by $softmax$ function.
%
%
At each step, predicted position $\bar{\mathbf{u}} = (\bar{u},\bar{v})$ and orientation $\bar{\alpha}$ of the agent are computed by weighted averages using weights $\{w_{p,i}\}_{i=1}^{N_{p}}$ as follows:

\vspace{-0.4cm}
\begin{align}
\begin{split}
    \hspace{-0.3em} \bar{\mathbf{u}} &= \frac{1}{N_{p}}\sum_{i=1}^{N_{p}}{w_{p,i}\mathbf{u}_{i}} \\
    \hspace{-0.3em} \bar{\mathbf{\alpha}} &= \atantwo{\left( 
    \frac{1}{N_{p}}\sum_{i=1}^{N_{p}}{w_{p,i}\sin{\alpha_{i}}},
    \frac{1}{N_{p}}\sum_{i=1}^{N_{p}}{w_{p,i}\cos{\alpha_{i}}}
    \right)}.
\end{split}
\end{align}
\vspace{-0.3cm}

\subsection{Navigation Using Jackal Robot}


\begin{figure}[t!]
{\centering
\includegraphics[width=1.0\linewidth]{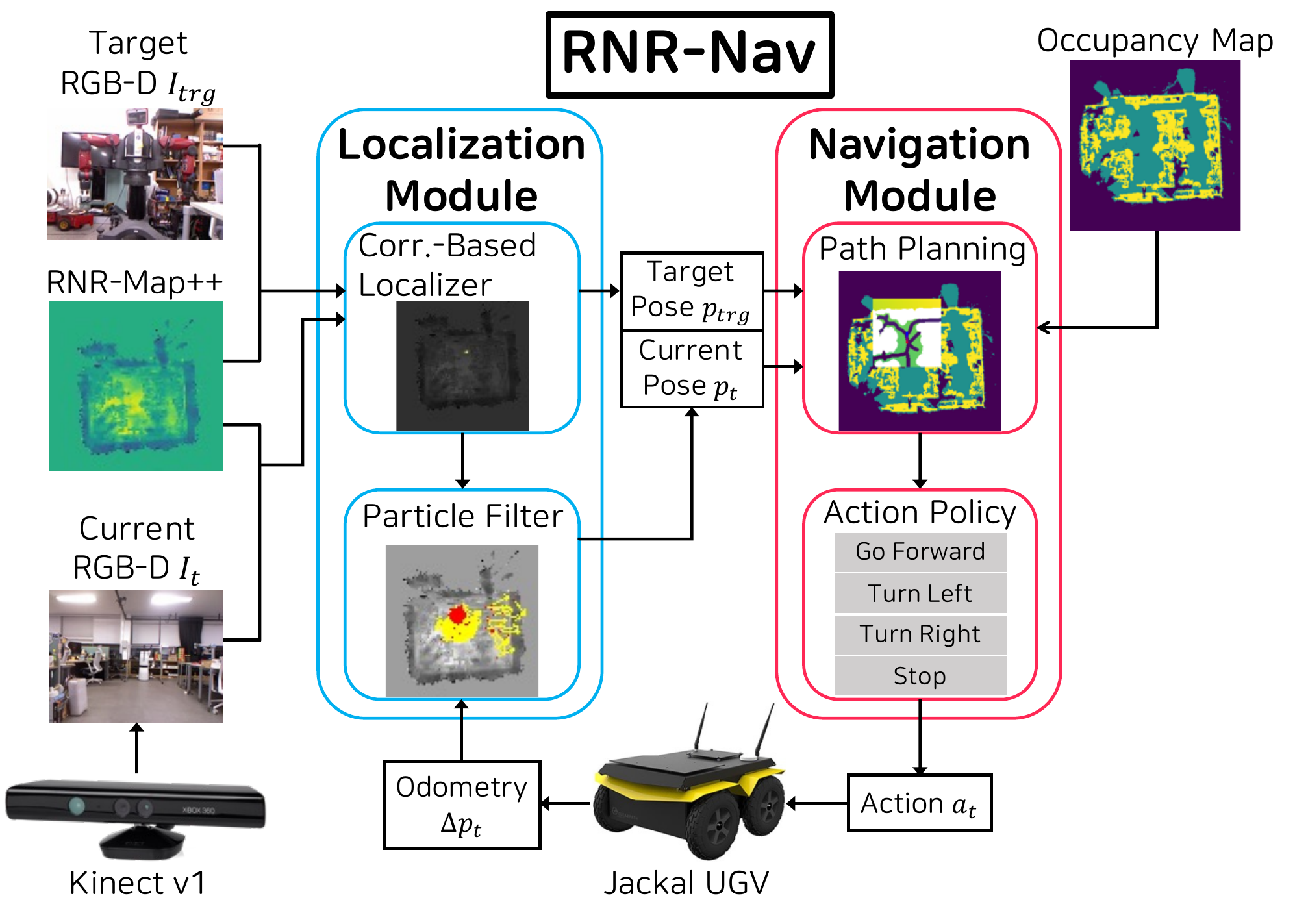}
}\centering
\vspace{-0.7cm}
\caption{\small \textbf{Overview of the proposed RNR-Nav navigation framework using RNR-Map++}. The localization module predicts the current pose from the current RGB-D image. The navigation module plans the path and gives the next action to the agent. An RNR-Map++ and an occupancy map need to be prepared.}
\label{fig:system}
\vspace{-0.5cm}
\end{figure}

%
We establish a real-world robot system using RNR-Map++ for image-goal navigation tasks, RNR-Nav.
We opt for the Jackal UGV as our robot platform and Kinect v1 camera to capture RGB-D images. Note that Jackal UGV provides estimated odometry values through its built-in sensors, which are suitable for the particle filter. Our RNR-Nav system comprises a localization module and a navigation module, as illustrated in Figure \ref{fig:system}. The workflow of the system is as follows: Prior to the navigation task, both an RNR-Map++ and an occupancy map of the environment should be prepared. At the beginning, the target pose $p_{trg}$ is determined based on the given target RGB-D image $I_{trg}$ utilizing the RNR-Map++ and correlation-based localizer. The localization module, consisting of a correlation-based localizer and a particle filter, keeps track of the pose of the robot. When it receives a new RGB-D observation $I_{t}$, the localizer predicts a heatmap and subsequently passes it to the particle filter. Simultaneously, the particle filter receives pose discrepancy information $\delta p_{t}$ from the odometry values of the Jackal robot.
%
Consequently, the particle filter provides a prediction regarding the current pose $p_{t}$ of the agent.

The navigation module determines the next action $a_{t}$ to be taken from the current pose. For safety, we employ a conservative approach by filtering out areas close to the obstacles based on the occupancy map. As the localization module provides the current pose, the planner sets a local goal and local path while ensuring adherence to the safe regions. When the robot reaches the local goal, or there is a significant change in the estimated pose compared to the previous one, the planner replans the local path. Subsequently, a suitable action $a_{t}$ is determined to facilitate the path following from the current position. The decision to stop is based on the relative distance between the estimated current position and the target position. We test our robot system in real-world environments, and the results are detailed in the following section.

\section{Experiments}

\subsection{Training Preparation}

For training, we use the ScanNet \cite{scannet} dataset, which consists of more than 1500 indoor sequences for training and 100 test sequences containing RGB-D observations and ground truth camera poses. We find that training with 500 of the 1500 sequences is sufficient for the generalizability.

We set the map size of RNR-Map++ to $(U,V)=(128,128)$ with grid size $s=0.25m$, which makes the map cover $32m \times 32m$ regions of the environment. To fulfill our constraint of camera parameters, $f_{x} = f_{y} = 128$, we use resized images with the shape of $(H,W)=(102,134)$ for ScanNet images. For our robot system with Kinect v1 camera, we use $(H,W)=(113,152)$ for input image size.
We measure and report the running times of various processes using a laptop PC with an Intel i7-10875H CPU @ 2.30GHz and an NVIDIA GeForce RTX 2070 Super GPU, which is also used for our real-world RNR-Nav experiment.




\subsection{Results}

\subsubsection{Reconstruction}

\begin{table}[tb!]
\caption{\small\textbf{Rendering quality comparisons of variants of RNR-Map on ScanNet \cite{scannet} test scenes.} The terms Weight and PE refer to whether using a weighted map and positional encoding for the encoder.}
\vspace{-0.3cm}
\begin{center}
\scalebox{0.8}{
\input{tables/recon_result}
}
\end{center}
\label{tab:recon_result}
\vspace{-0.3cm}
\end{table}

\begin{table}[tb!]

\caption{\small\textbf{Localization results on ScanNet \cite{scannet} test scenes.} The term PF refers to a particle filter, and the term RR refers to a robustness ratio.}
\vspace{-0.3cm}
\begin{center}
\scalebox{0.9}{
\input{tables/loc_result}
}
\end{center}
\label{tab:loc_result}
\vspace{-0.6cm}
\end{table}

We evaluate the rendering performance on the ScanNet test dataset, which consists of 100 scenes and more than 200k RGB-D images with known poses. Note that they are unseen frames during training.
The entire map is constructed beforehand and subsequently used to render images at specified poses.
The mapping process $F_{reg}$ processes each image of the ScanNet dataset in just $12.1ms$, enabling operation at a speed of 82.6Hz.
We only use uniformly sampled 10\% of the total frames to make the maps. For rendering, we test with two subsets: \textit{Seen} - about 20k frames that are used to construct maps, and \textit{Full} - about 200k full frames that include seen and unseen images.

We measure and report the three widely used metrics representing image quality: PSNR, SSIM, and LPIPS. The results shown in Table \ref{tab:recon_result} confirm that the reconstruction performance remains consistent regardless of whether the frame is observed before. It verifies that the RNR-Map++ is robust in terms of the number of images for construction.
The results indicate that incorporating positional encoding (PE) has a notable impact on improving rendering performance. It suggests that maintaining vertical positional information is essential for creating visually informative maps. Additionally, the weighted map demonstrates quantitative improvements, suggesting that capturing salient features from images onto the map yields more effective results.
%

As we can see from the quantitative result and qualitative result from Figure \ref{fig:recon_quality}, the RNR-Map++ shows robustness for unseen environments, such as the ScanNet test set or our own real-world observations. Unlike several NeRF-based methods that are limited to use in a single scene, our approach is applicable to any environment after training once.

\subsubsection{Localization}

With the trained encoder, we could make an egocentric query RNR-Map++ from a query RGB-D image. Constructed RNR-Map++ of test scenes are used as key maps. We evaluate localization performance by estimating the poses of the 200k \textit{Full} subset for $R=18$ angle bins. From the ground truth poses, we compute the relative pose differences for the particle filter with additional noises.

We compare our method to the localization method of Loc-NeRF \cite{locnerf}, which uses rendering processes and a particle filter. Since the scales of ScanNet scenes are more extensive than \cite{locnerf} settings, and the performance inevitably depends on the number of particles, we adjust the initial number of particles proportional to the volume of each scene with a maximum limit of 500. In addition, we measure the performance of the localization process using the original version of RNR-Map and its correlation-based localizer.

We use the following metrics for the localization task: 1) average distance error $e_{dist}$, 2) average absolute value of orientation error $e_{ori}$, and 3) robustness rate $RR$, which denotes the ratio of when $e_{dist}$ is smaller than $0.5m$. Table \ref{tab:loc_result} provides a comprehensive overview of the results. Notably, the particle filter shows significant improvements in both accuracy and robustness. Specifically, in the case of RNR-Map++, we observe a substantial 59.4\% reduction in the $e_{dist}$ value. The high robustness rate of 97.1\% indicates the applicability of the proposed localization process in diverse tasks such as navigation.
Moreover, the results reveal that effectively integrating visual information into RNR-Map++ is advantageous for correlation-based localization compared to RNR-Map.
%
In contrast, the Loc-NeRF exhibits less favorable results with a robustness rate of only 44.3\%. We find that its performance is susceptible to the convergence of particles in the early stages, often displaying high levels of randomness.
%
The proposed localization method using RNR-Map++ with particle filter shows an average inference time of $33.8ms$ (29.6Hz), confirming it is reasonable to consider as a real-time process.
On the contrary, Loc-NeRF requires an average of $2340ms$ (0.43Hz) per frame, which makes it challenging to serve as a real-time process.
%

\subsection{Jackal Demonstration}

\begin{figure}[t!]{\centering\includegraphics[width=0.83\linewidth]{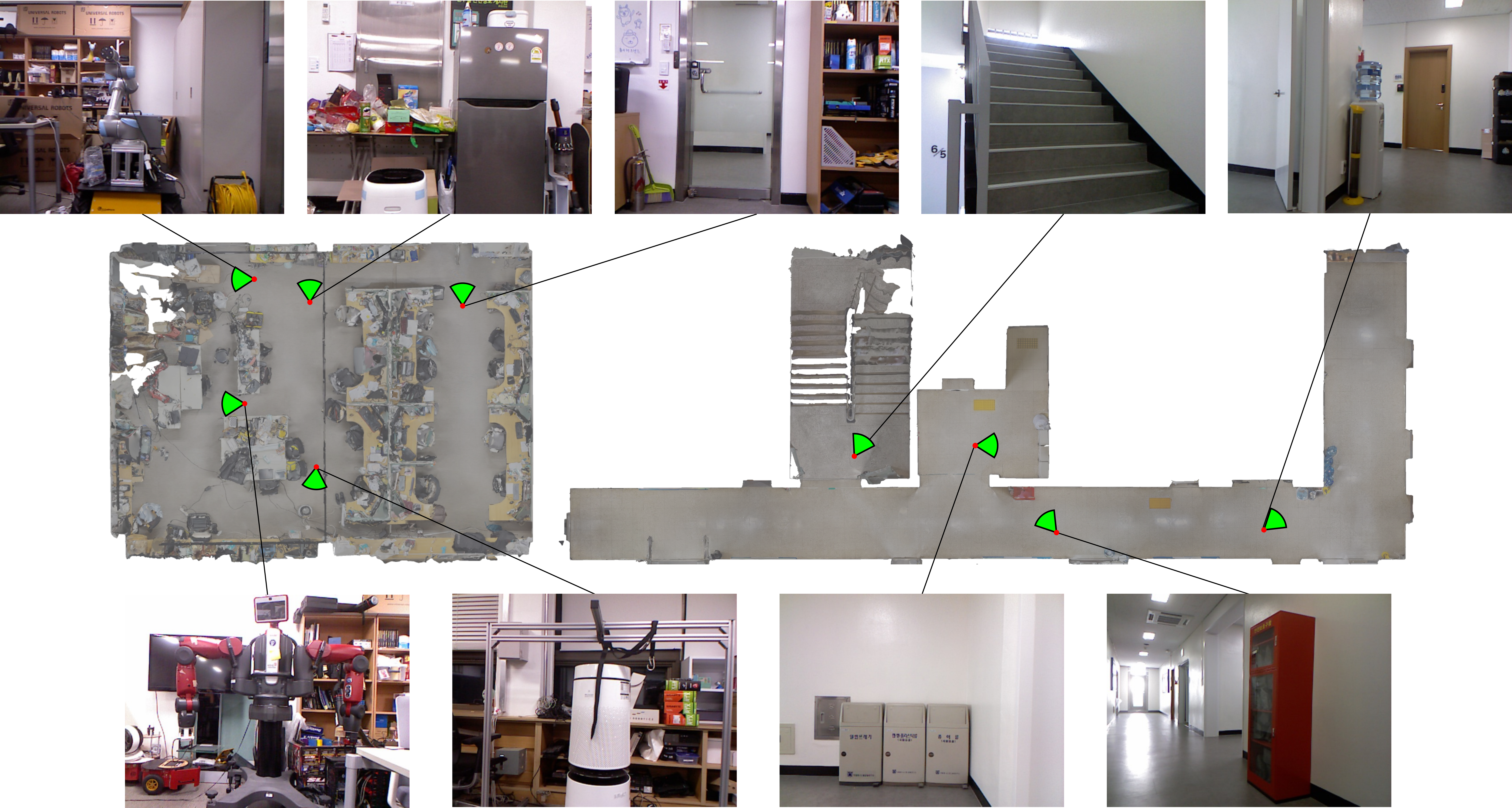}}\centering
\vspace{-0.1cm}
\caption{\small\textbf{Real-world environments and various landmarks.} We set two environments: \textit{Laboratory} and \textit{Corridor}. Landmarks with distinct views are selected for starting points and target points.}
\label{fig:query}
\vspace{-0.5cm}
\end{figure}

The visual navigation experiment is conducted in two real-world environments: \textit{Laboratory} and \textit{Corridor}. We select characteristic landmarks, five for the laboratory and four for the corridor, as shown in Figure \ref{fig:query}. They pair up with each other as starting points and goal points, which gives 20 pairs of trajectories for the laboratory and 12 for the corridor. Before navigation, RNR-Map++ and the occupancy map are required. So, we first drive the Jackal robot around two environments thoroughly and record RGB-D images. After calibrating the intrinsic parameters of the Kinect v1 RGB-D camera, we gain estimated poses of the images using ORB-SLAM3 \cite{orb_slam3}, which we treat as ground truth poses. With ground truth poses, we construct RNR-Map++ and the occupancy map.

We test navigation performance on 32 trajectories of the proposed RNR-Nav and the original RNR-Map navigation system. For both cases, the correlation-based localizer estimates the starting point and the goal point. For RNR-Map, rendering-based localization is used to refine the estimated pose. 
As an output of the navigation module, we implement to return one of the four discrete actions: to move forward for $25cm$, to turn left for $10^\circ$, to turn right for $10^\circ$, and to stop when the distance to the goal is lower than $30cm$.

\begin{table}[tb!]

\caption{\small\textbf{Navigation results in real-world scenarios.} The term Optim. refers to optimization-based localization.}
\vspace{-0.2cm}
\begin{center}

\scalebox{0.85}{
\input{tables/jackal_result}
}
\end{center}
\label{tab:jackal_result}
\vspace{-0.6cm}
\end{table}

The results are shown in Table \ref{tab:jackal_result}. RNR-Nav demonstrates a high success rate in navigation within the laboratory environment, which has sufficient visually distinct objects. It facilitates the correlation-based localizer to predict the pose of the robot without confusion. The sole failure has arisen from a collision in a narrow pathway. Navigation through the corridor appears more challenging due to the prevalence of similar monotonous observations like empty walls. Nonetheless, during successful episodes, the particle filter prove invaluable, effectively refining the pose whenever the visual localizer faced uncertainties.

In contrast, the navigation framework employing the original RNR-Map and optimization-based localization exhibits a notably low success rate compared to RNR-Nav. Since the performance of the correlation-based localizer falls short of the proposed method, it often mislocates the starting points or the goal points. An incorrect prediction of the goal pose inevitably leads to a completely different trajectory. Rectifying an erroneous starting point is only possible through an optimization process, which is exceedingly challenging with limited iterations.
Even when the initial and final points of the trajectory are accurately predicted, the optimizer demonstrates inaccurate outcomes since the performance depends on the rendering quality.

We also measure the speed of the localization process during navigation. The proposed localization method with the particle filter demonstrates an average execution time of $39.4ms$ (25.4Hz), which is slightly longer than the ScanNet case because of the bigger image size. It is more than six times faster than the running time of $248ms$ (4.03Hz) using rendering and optimization.


\section{Conclusion}
In this paper, we have proposed a novel real-world image-goal navigation framework, RNR-Nav, based on RNR-Map++.
RNR-Map++ adopts weighted map and positional encoding to reduce information loss during map construction. In addition, we have introduced a robust localization method utilizing cross-correlation of RNR-Map++ and particle filter. The experimental results show enhanced performance of RNR-Map++ and the localization method. Moreover, RNR-Nav demonstration in real-world environments achieves a high success rate.

{\small
\bibliographystyle{IEEEtran}
\bibliography{IEEEabrv, ref}
}

\end{document}

%% file: abstract/abs2.tex
We propose a novel visual localization and navigation framework for real-world environments directly integrating observed visual information into the bird-eye-view map.
%
While the renderable neural radiance map (RNR-Map) \cite{RNR-Map} shows considerable promise in simulated settings, its deployment in real-world scenarios poses undiscovered challenges.
RNR-Map utilizes projections of multiple vectors into a single latent code, resulting in information loss under suboptimal conditions.
To address such issues, our enhanced RNR-Map for real-world robots, RNR-Map++, incorporates strategies to mitigate information loss, such as a weighted map and positional encoding.
For robust real-time localization, we integrate a particle filter into the correlation-based localization framework using RNR-Map++ without a rendering procedure.
Consequently, we establish a real-world robot system for visual navigation utilizing RNR-Map++, which we call ``RNR-Nav."
Experimental results demonstrate that the proposed methods significantly enhance rendering quality and localization robustness compared to previous approaches.
In real-world navigation tasks, RNR-Nav achieves a success rate of 84.4\%, marking a 68.8\% enhancement over the methods of the original RNR-Map paper.

%% file: figures/weight.tex
\begin{tabular}{cc}
\begin{tabular}{@{\hspace{-0.1cm}}c@{\hspace{-0.5cm}}}\small Image \end{tabular} &
\begin{tabular}{c} \includegraphics[width=0.75\linewidth]{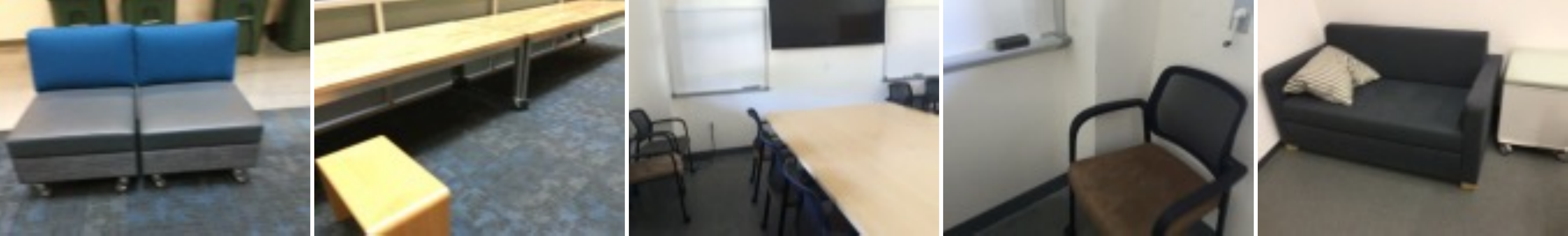} \end{tabular}\vspace{-0.05cm}\\
\begin{tabular}{@{\hspace{-0.1cm}}c@{\hspace{-0.5cm}}} \small Weight \end{tabular} &
\begin{tabular}{c} \includegraphics[width=0.75\linewidth]{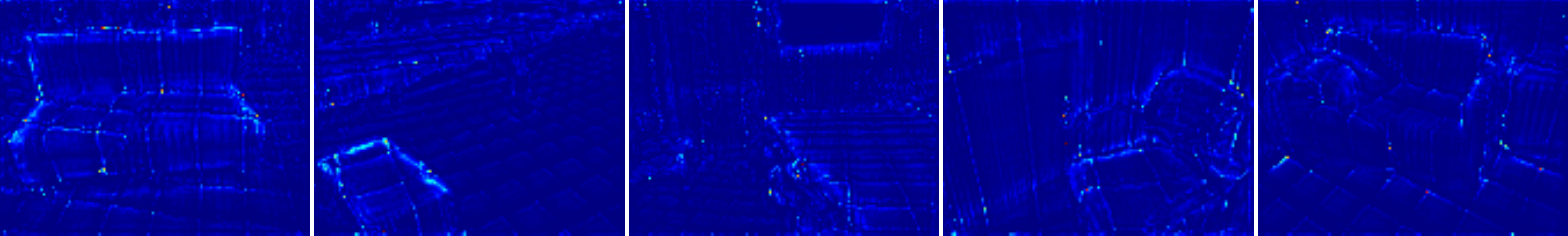} \end{tabular}

\end{tabular}

%% file: figures/recon_quality.tex
\begin{tabular}{cc}
\begin{tabular}{@{\hspace{-0.1cm}}c@{\hspace{-0.5cm}}}\small G.T. \end{tabular} &
\begin{tabular}{c} \includegraphics[width=0.75\linewidth]{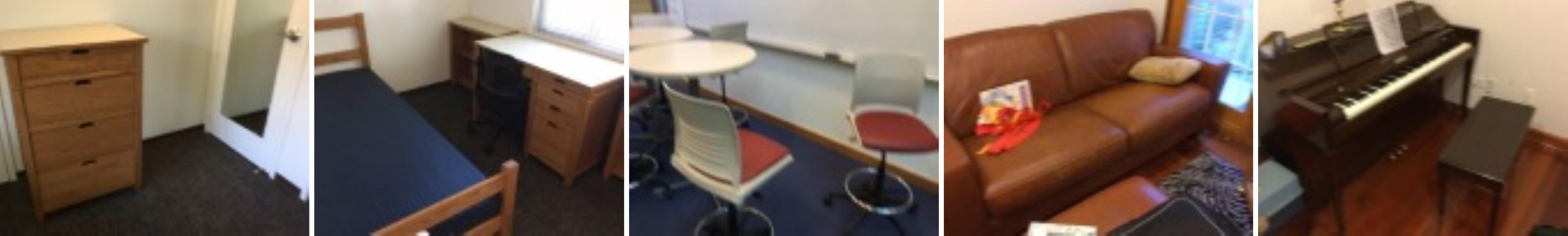} \end{tabular} \vspace{-0.05cm}\\
\begin{tabular}{@{\hspace{-0.1cm}}c@{\hspace{-0.5cm}}}\small RNR-Map \end{tabular} &
\begin{tabular}{c} \includegraphics[width=0.75\linewidth]{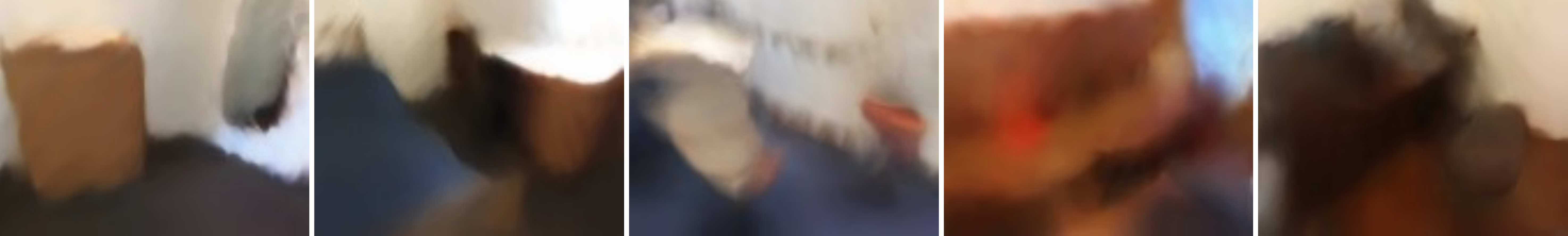} \end{tabular} \vspace{-0.05cm}\\
\begin{tabular}{@{\hspace{-0.1cm}}c@{\hspace{-0.5cm}}}\small RNR-Map++ \end{tabular} &
\begin{tabular}{c} \includegraphics[width=0.75\linewidth]{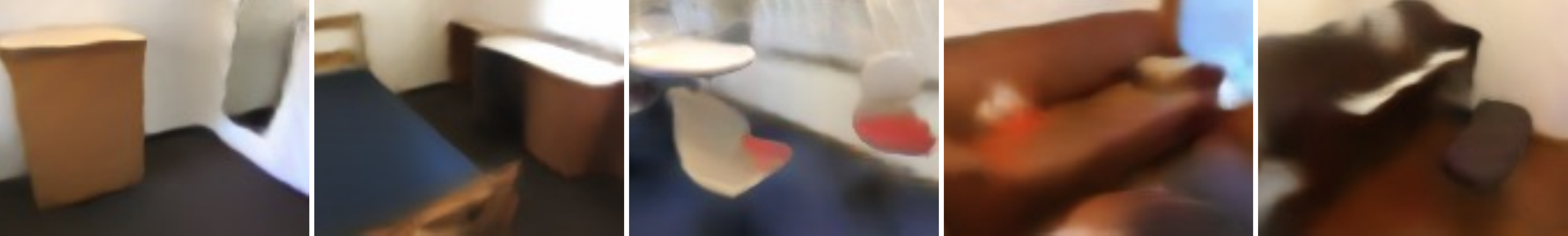} \end{tabular} \vspace{-0.1cm}\\
& \vspace{0.1cm} \begin{tabular}{c} (a) \textbf{\small ScanNet \cite{scannet} Dataset} \end{tabular}\\
\begin{tabular}{@{\hspace{-0.1cm}}c@{\hspace{-0.5cm}}}\small G.T. \end{tabular} &
\begin{tabular}{c} \includegraphics[width=0.75\linewidth]{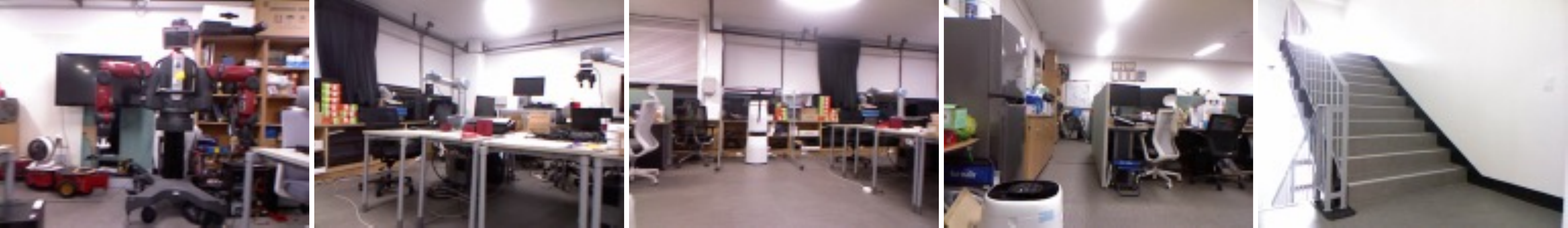} \end{tabular} \vspace{-0.05cm}\\
\begin{tabular}{@{\hspace{-0.1cm}}c@{\hspace{-0.5cm}}}\small RNR-Map \end{tabular} &
\begin{tabular}{c} \includegraphics[width=0.75\linewidth]{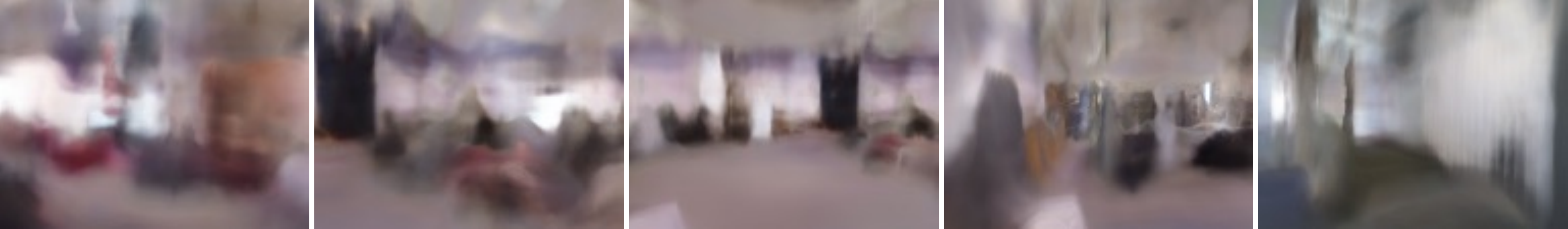} \end{tabular} \vspace{-0.05cm}\\
\begin{tabular}{@{\hspace{-0.1cm}}c@{\hspace{-0.5cm}}}\small RNR-Map++ \end{tabular} &
\begin{tabular}{c} \includegraphics[width=0.75\linewidth]{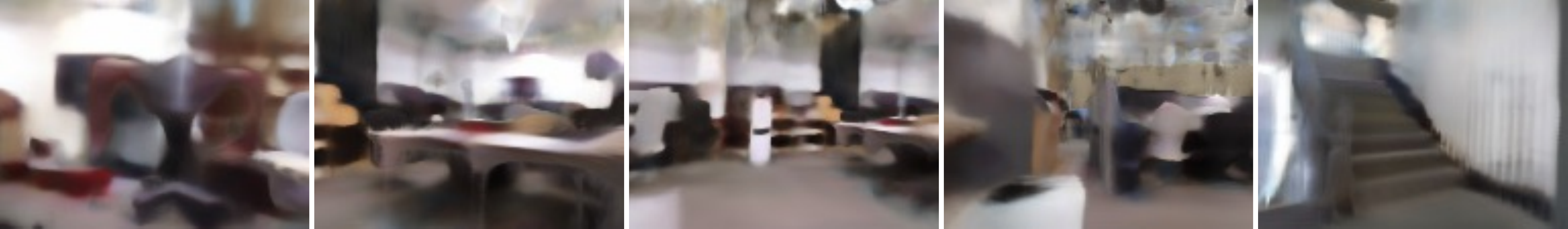} \end{tabular} \vspace{-0.1cm}\\
& \vspace{0.1cm} \begin{tabular}{c} (b) \textbf{\small Real-World Observation} \end{tabular} \\
\end{tabular}

%% file: tables/recon_result.tex
\begin{tabular}{cc|cccccc}

\toprule
\multicolumn{2}{c|}{} & \multicolumn{3}{c}{\textbf{Seen}} & \multicolumn{3}{c}{\textbf{Full}\textbf} \\
\cmidrule(lr){3-5}\cmidrule(lr){6-8}
Weight & PE & PSNR $\uparrow$ & SSIM $\uparrow$ & LPIPS $\downarrow$ & PSNR $\uparrow$ & SSIM $\uparrow$ & LPIPS $\downarrow$ \\
\midrule
\xmark & \xmark & 16.318 & 0.465 & 0.613 & 16.315 & 0.466 & 0.613 \\
\checkmark & \xmark & 16.351 & 0.468 & 0.611 & 16.607 & 0.470 & 0.610 \\
\xmark & \checkmark & 18.972 & 0.566 & 0.536 & 18.965 & 0.566 & 0.536 \\
\midrule
\checkmark & \checkmark & \textbf{19.113} & \textbf{0.571} & \textbf{0.533} & \textbf{19.106} & \textbf{0.571} & \textbf{0.532} \\
\bottomrule

\end{tabular}

%% file: tables/loc_result.tex
\begin{tabular}{l|ccc}

\toprule

\textbf{Method} & $\boldsymbol{e_{dist}(m)}$ $\downarrow$ & $\boldsymbol{e_{ori}(^{\circ})}$ $\downarrow$ & \textbf{RR(\%)} $\uparrow$ \\

\midrule

Loc-NeRF \cite{locnerf} & 1.472 & {44.738} & 44.3 \\

\midrule

RNR-Map \cite{RNR-Map} & 0.375 & {14.671} & 90.1 \\
RNR-Map w/ PF & 0.156 & { 5.804} & 95.7 \\

\midrule

RNR-Map++ & 0.293 & {10.552} & 94.5 \\
\textbf{RNR-Map++ w/ PF} & \textbf{0.119} & \textbf{ 4.117} & \textbf{97.1} \\

\bottomrule

\end{tabular}

%% file: tables/jackal_result.tex
\begin{tabular}{@{}c c | c c@{}}
\toprule
\textbf{Method}                        & \textbf{Environment} & {\textbf{Success Rate} (\%) $\uparrow$} & \textbf{Loc. Time($\boldsymbol{ms}$)} $\downarrow$ \\ \midrule
\multirow{3}{*}{\begin{tabular}[c]{@{}c@{}}RNR-Map \cite{RNR-Map}\\ w/ Optim.\end{tabular}}
& Laboratory & {20.0 ( 4/20)} & \multirow{3}{*}{\begin{tabular}[c]{@{}c@{}}248 \\ (4.03Hz)\end{tabular}} \\
& Corridor   & { 8.3 ( 1/12)} & \\
& Total      & {15.6 ( 5/32)} & \\ \midrule
\multirow{3}{*}{\textbf{RNR-Nav (Ours)}}
& Laboratory & {95.0 (19/20)} & \multirow{3}{*}{\begin{tabular}[c]{@{}c@{}}\textbf{39.4} \\ \textbf{(25.4Hz)}\end{tabular}} \\
& Corridor   & {66.7 ( 8/12)}  & \\
& Total      & \textbf{84.4 (27/32)} & \\ \bottomrule
\end{tabular}